\documentclass[10pt, conference, letterpaper]{IEEEtran}

\usepackage{url}
\usepackage{graphicx,epsfig,subfig}
\usepackage{graphics,dcolumn,bm,epic,eepic,fleqn,float}
\usepackage{epstopdf}
\usepackage{amssymb,amsmath,multirow,rotate,color}
\usepackage[draft]{hyperref}
\usepackage{morefloats}
\usepackage{subfig}
\usepackage[draft]{hyperref}
\usepackage[T1]{fontenc}
\usepackage{color}

\usepackage{balance}
\usepackage{verbatim}

\usepackage{times}

\newcommand\fref[1]{Figure~\ref{#1}}
\newcommand\reffig[1]{\fref{#1}}

\begin{document}

\title{On Factors Affecting the Usage and Adoption of a Nation-wide TV Streaming Service}

\author{\IEEEauthorblockN{Dmytro Karamshuk\IEEEauthorrefmark{1},
Nishanth Sastry\IEEEauthorrefmark{1},
Andrew Secker\IEEEauthorrefmark{2} and 
Jigna Chandaria\IEEEauthorrefmark{2}}
\IEEEauthorblockA{\IEEEauthorrefmark{1}Department of Informatics, King's College London, UK. 
Email: {firstname.lastname}@kcl.ac.uk}
\IEEEauthorblockA{\IEEEauthorrefmark{2}BBC R\&D, London, UK. Email: {firstname.lastname}@rd.bbc.co.uk}
}

\maketitle 


\begin{abstract}
Using nine months of access logs comprising 1.9
Billion sessions to BBC iPlayer, we survey the UK ISP ecosystem
to understand the factors affecting adoption and usage of a high bandwidth
TV streaming application across different providers.
We find evidence that connection speeds are important and that
external events can have a huge impact for live TV usage. Then,
through a temporal analysis of the access logs, we demonstrate
that data usage caps imposed by mobile ISPs significantly affect
usage patterns, and look for solutions. We show that product
bundle discounts with a related fixed-line ISP, a strategy already
employed by some mobile providers, can better support user
needs and capture a bigger share of accesses. We observe that
users regularly split their sessions between mobile and fixed-line
connections, suggesting a straightforward strategy for offloading
by speculatively pre-fetching content from a fixed-line ISP before
access on mobile devices.
\end{abstract}

\section{Introduction}
\label{sec:intro}
As Internet usage matures and ``broadband'' access becomes near  ubiquitous, the variety of access network  choices available to users has increased. Beyond the traditional DSL and cable-based broadband providers, users can now also obtain data connectivity from mobile or cellular network operators. With the roll-out of 4G/LTE technologies, we are now at an inflection point where  achievable data rates from cellular connectivity can, in theory, exceed the rates of a majority of existing fixed-line broadband connections. 

However, this choice in Internet access also means that consumers face a complex array of choices with different providers offering different data access rates, amounts of data allowed, pricing structures as well as  auxilliary benefits such as discounts for product bundles (e.g., so-called ``triple'' or ``double'' play for cable TV access, or mobile providers who may offer discounted home broadband packages). 

This paper seeks to obtain a preliminary understanding of how differences in Internet access options affect consumer choice and usage behaviours for high bandwidth applications such as on-demand streaming. 
In particular, given the increasing importance of mobile access, 
we wish to understand the relative roles and importance of broadband and cellular network access for users across the nation. 

We study these questions from a \emph{first-mile viewpoint} with an unusually broad view of the UK Internet that makes it suitable to obtain an understanding of the nationwide ISP ecosystem: BBC iPlayer is a highly popular ``catch-up'' TV application that allows over-the-top streaming access to TV and radio content recently broadcast by the BBC. In 2012, Ofcom, UK's communications regulator, estimated that over 44\% of UK households had accessed iPlayer~\cite{ofcom}, and Sandvine estimated it to be the most popular streaming application on UK's networks, after YouTube~\cite{sandvine}; access numbers have increased significantly since 2012~\cite{bbc_performance_pack_2014_march}. 

We obtain access logs to BBC iPlayer for a nine month-long period from May 2013-Jan 2014. Our trace covers 1.9 billion sessions of 32M monthly users (corresponding to 50\% of UK's population) from 20M monthly IP addresses. 
Uniquely, because the iPlayer is funded by TV Licensing Fees paid by UK residents, accesses to TV content, which comprise over 75\% of sessions, is restricted to the UK. Thus, these session numbers are driven primarily by UK consumers, which makes it convenient to study the UK market.  In particular, all the major UK providers, both fixed-line broadband and cellular operators, are well represented. 



We first study support for the basic needs of TV streaming: the need for high bandwidth, the need for supporting traffic spikes, and the need to support differences in usage patterns of mobile devices and traditional computing equipment (laptops, Desktop PCs) as compared with Traditional TV (\S\ref{sec:pop-view}).
Then, we exploit the long duration of our trace to ask how the ISP market evolves over time, and uncover the temporal characteristics of mobile vs.\ broadband access  (\S\ref{sec:over-time}).  Finally, in \S\ref{sec:crossisp}, we study how users' sessions are split across different providers and show how this is affected by product pricing and bundling strategies that provide discounts for combinations of fixed and mobile ISPs. We suggest ways to exploit consumers' usage patterns across mobile and fixed ISPs to shift traffic from mobile to fixed ISPs using speculative pre-fetching.


Our major findings may be summarised as follows:
\begin{enumerate}
\item We demonstrate that iPlayer usage in a geographic region is correlated with average broadband speeds in that region, underscoring the need for high-speed broadband infrastructure.

\item We find that mobile devices have a disproportionate share of live accesses, and external events such as Wimbledon can have a significant impact.

\item Distinct diurnal patterns of access are seen for different kinds of providers: Fixed-line broadband peaks during evening hours whereas cellular network access peaks during commute times, for mobile providers with limited data plans. Mobile providers with unlimited plans observe a \emph{superposition} of both patterns, with peaks during commutes and evening hours.
\item We characterise the load shares of users' iPlayer sessions across different providers and find, as expected that fixed-line broadband captures the highest fraction of users' session loads. We also show that offering unlimited data plans captures significantly higher fraction of users' iPlayer sessions for mobile providers. However, we also find that an alternate strategy of offering price discounts and product bundles with a dedicated fixed-line broadband infrastructure can also serve to capture greater shares of users' sessions and make both the fixed-line and mobile provider attractive as a package to customers.
\item Mobile users often split their content consumption across different sessions. Such sessions are either first started on fixed-line broadband and finished while on the move (53\%), or started on a cellular connection and finished later on a fixed connection (47\%). In particular, 25\% of the 47\% sessions started first on mobile networks are for new episodes of a regularly watched TV show. 
We argue that such simple viewing patterns offer significant opportunities for pre-fetching content from a fixed connection and offload data from mobile broadband connections.
\item Users over mobile connections access more adult and less children-related content than over fixed-line connections. Because children-related content can be up to 25\% on fixed-line connections, this has important implications for caching and cache-peering across different kinds of providers. 
\end{enumerate}

\section{Related Work}

As the Internet and the Web have grown in importance, there have been numerous attempts in the past decade to study end-system behaviours and Internet Service Provider characteristics. Perhaps the most straightforward approach, followed by a number of studies~\cite{lakshminarayanan2003some,netiathome,dischinger2007characterizing,kreibich2010netalyzr,sundaresan2011broadband,sanchez2013dasu}, is to directly recruit end hosts to run measurements. Although this \emph{last-mile viewpoint} has shed light on significant access network bottlenecks as well as a wide array of end-user and access network properties, the logistics of obtaining multiple vantage points at the edge remains a challenge; thus the scale of these efforts has been limited, ranging from a few tens~\cite{lakshminarayanan2003some}, hundreds~\cite{netiathome} or thousands of hosts~\cite{dischinger2007characterizing,sundaresan2011broadband}, to slightly less than one hundred thousand hosts~\cite{kreibich2010netalyzr,sanchez2013dasu} for the largest studies to date. 

A second set of studies  has focused on the middle, examining data obtained from ISPs, whether in the core (e.g.,~\cite{xu2005profiling}), or in access networks close to consumers (e.g.,~\cite{karagiannis2005blinc,moore2005toward,maier2009dominant}), to obtain an aggregate picture of Internet traffic. The more central ISP-based viewpoint of such studies has provided valuable information about end user behaviours, especially the changing mix of applications on the Internet,  the relative proportions and popularities of different types of traffic such as P2P and HTTP, etc. However, because of the difficulty of obtaining ISP data, previous academic efforts using this approach have mostly relied on data from one or two ISPs. While these efforts are now replicated at a global scale by commercial vendors such as Sandvine in their biannual Internet Performance Report~\cite{sandvine}, coverage is still selective,  based on the 250 or so fixed and mobile operators who are customers of Sandvine.

We adopt a \emph{first-mile} approach -- observing end hosts through the perspective of a server or content provider. This viewpoint is complementary to the above efforts: We only have one application in our traffic mix -- BBC iPlayer provides over-the-top streaming of TV and radio content that has been recently broadcast by the BBC. However, as explained in \S\ref{sec:intro}, BBC iPlayer is extremely popular within the United Kingdom, which allows us to get a good picture of a nation's Internet service ecosystem.  
Not many have followed the first mile approach. In~\cite{li2012watching} Liu et. al have looked at a similar dataset from a large video content provider in China. However, we consider a significantly longer time span of accesses (9 months in comparison to 2 weeks), which captures long-term patterns in the behavior of the national consumer (\S\ref{sec:9monthview}).


Characterisation and measurement of application- and IP Network-level performance of 3G/4G has been more recent. \cite{xu2011cellular} characterise the infrastructure of the four major cellular carriers in the US using a location search app. The authors identify potential placement strategies for mobile content and find that too few gateways in cellular providers mean that content delivery is difficult for them. \cite{huang2013depth} studies TCP performance on LTE and identifies potential bottlenecks, whereas \cite{zhu2011characterizing} analyses diurnal patterns of users in 3G data networks. With respect to this stream of work, we focus on a single application---streaming, which allows us to identify distinct usage patterns across different types of networks. 

One of the contributions of this work is a study covering what we believe to be a large fraction of the IP space of a country. \cite{heidemann2008census} conduct a thorough census of the Internet. In contrast to our approach, this doesn't capture application-layer and has only IP-level details. Similarly, Akamai regularly reports on the current state of the Internet, comparing the relative market shares of different devices, and average broadband rates obtained, and recent reports~\cite{akamai-soti} have included basic information about mobile connectivity. However, it is unclear whether they can identify users across different networks; this is critical, for example, to analyse the impact of a mobile connection on fixed-line broadband and vice versa (Figure~\ref{fig:cross_isp_ties}). 

\section{Dataset}

\subsection{Introduction to BBC iPlayer}
The BBC has a number of local and national TV and radio channels, which broadcast content over the air in the UK. The BBC iPlayer is an Internet streaming service that makes this broadcast content  available for streaming on the iPlayer website. Content is typically made available for a fixed period of days after the broadcast, depending on content licensing terms and other policies. Thus, the iPlayer provides an alternate ``over-the-top'' access mechanism for content  broadcast over the air. iPlayer additionally provides live streaming access to content currently being broadcast, but accessing a live video stream requires a TV license to be paid for annually. After  broadcast, most content is made available for ``catch up'' viewing, and may be accessed even without a TV license by viewers within UK. Currently, on-demand access constitutes the vast majority ($\approx$90\%) of TV sessions. 

\subsection{Dataset description}
In this paper we consider a nine month snapshot of access logs to BBC iPlayer in May, 2013 - January, 2014. The access logs have been collected by aggregating over "heartbeats"
generated regularly (each 10-30 seconds) or on a user action (e.g., video window resize or pause) between iPlayer client software and a statistics server. All data have been anonymised before analysis. Each record in the dataset contains information about a user's session in the following format: 
\newline

\emph{$<$network-id, isp, geographic region, user-id, content-id, content genre, session start time, duration, device type$>$}. \newline

The anonymized \emph{network-id} identifies each unique IP address seen in our trace separately. The \emph{ISP} and \emph{geo region} corresponding to this network-id/IP address are resolved before anonymization using the RIPE\footnote{http://www.ripe.net/data-tools/db} and MaxMind\footnote{http://dev.maxmind.com/geoip/legacy/geolite/} databases, respectively. Multiple ASes belonging to the same ISP or cellular network have been manually identified and merged together for the providers we consider. This also takes into account various recent mergers and acquisitions that are not yet reflected in RIPE. Overall, for ISP-related analysis we consider the five largest fixed-line and the five largest mobile ISPs which together account for $\approx 70$\% of users' sessions. One limitation is that geo-location information for cellular networks is known to be highly inaccurate~\cite{triukose2012geolocating}. Therefore, we only consider the five biggest fixed-line ISPs when using geo-location information.

\begin{table}
\small{
\centering
\begin{tabular}{r|c|c}
& \textbf{All} & \textbf{January} \\
\hline
Number of Users & 156 M  & 32 M\\
Number of IP addresses & 55 M & 20 M\\
Number of Sessions & 1,951 M & 265 M \\
\hline
\end{tabular}
\caption{The dataset of accesses to BBC iPlayer in May, 2013 - January, 2014}\vspace{-6mm}
\label{tbl:dataset}}
\end{table}

The anonymized \emph{user-id} is based on long-term cookies (with a four year expiration date), that uniquely identifies each user agent separately. A single user might have more than one \emph{user-id} if they use more than one device, or even if they use more than one browser to access iPlayer. Users might also get multiple IDs if their cookies expire. However, in this paper we refer to users as identified by the user-ids. 


The other fields contain various information about the session, such as the session start time, its duration\footnote{The session duration accounts for number of seconds of the content item watched by user, buffering and pauses are not included.}, the content being watched and the device type. The type of device is extracted by matching user-agent HTTP headers of a user's request over 18 different string patterns (i.e., which together account for over $96$\% of iPlayer sessions) corresponding to the most popular devices (e.g., iphone, android gadget, macintosh computer, playstation) in one of the three groups: mobile gadgets (smartphones and tablets), PC-like devices (desktop computers and laptops), and Internet-enabled TVs (game consoles and TV set-top boxes). The content-id field indicates a unique id of a TV or radio show for on-demand sessions (e.g., "Sherlock Holmes, Series 1, Episode 2") or channel name for live accesses (e.g., BBC One, BBC Four). Additionally, the content genre field indicates one (or few) out of 11 iPlayer categories (e.g., drama, comedy, children) to which a content item belongs. Table~\ref{tbl:dataset} gives an indication of the size of the data. 


\begin{figure}[!tbp]
     \centering
     \includegraphics[width=0.7\columnwidth]{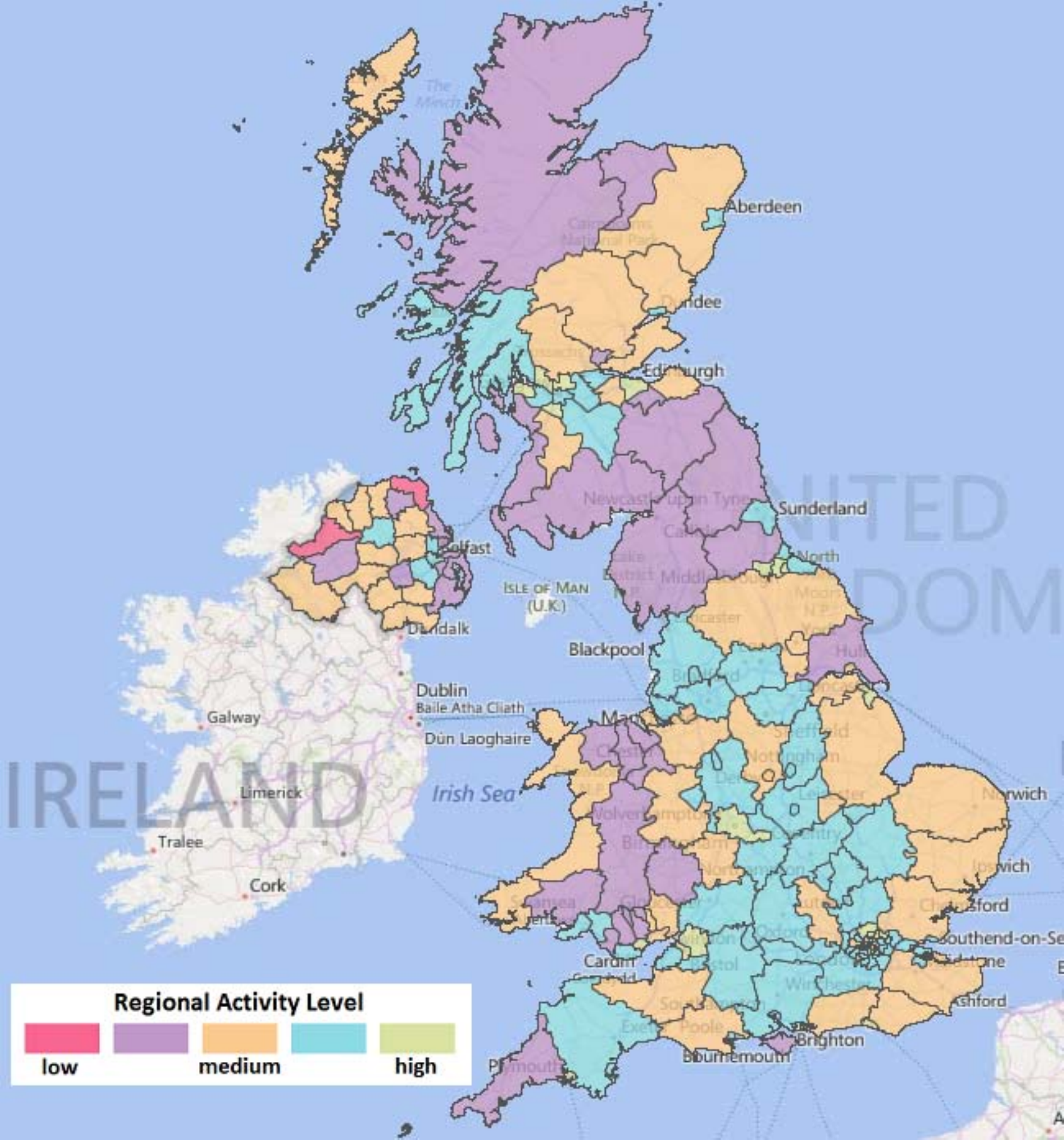}
     \caption{Variations in iPlayer's regional activity levels across the UK. The map shows clear differences across geographic regions where user activity varies from extremely heavy (olive green) to extremely low (pink). Only users of fixed-line internet are taken into account.}\vspace{-6mm}
     \label{fig:activity_per_region}
   \end{figure}
	
	\begin{figure}[!tbp]
     \centering
     \includegraphics[width=0.7\columnwidth]{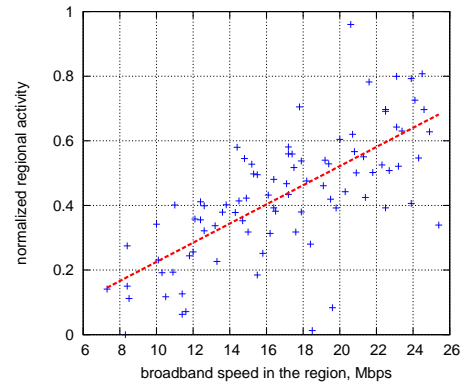}
     \caption{Regional activity of BBC iPlayer users is highly correlated (Pearson's correlation coefficient of 0.68) with the average broadband connection speed available in the region~\cite{ofcom_geography}.}\vspace{-6mm}
     \label{fig:activity_per_region_stats}
\end{figure}

\section{Population-level view of BBC iPlayer usage}
\label{sec:pop-view}
BBC iPlayer is an extremely popular application in the UK, used by an estimated 44\% of the population~\cite{ofcom}. However, different users use it to different extents. In this section, we seek to understand these differences, based on where (i.e., which geographic region) the users come from, and the kind of devices they use to access iPlayer. The heavy (and light) users are distributed across all the major ISPs. We defer a discussion of differences by ISP-type (mobile vs.\ fixed-line broadband) until the next section.  
 We use  number of sessions per user as a proxy for how heavy the usage or activity level of a user is. 

\begin{figure*}[htbp]
  \centering
  \subfloat[CDF of user activity levels]  {
    \label{fig:activity-level-gadgets}
    \includegraphics[width=0.32\textwidth]{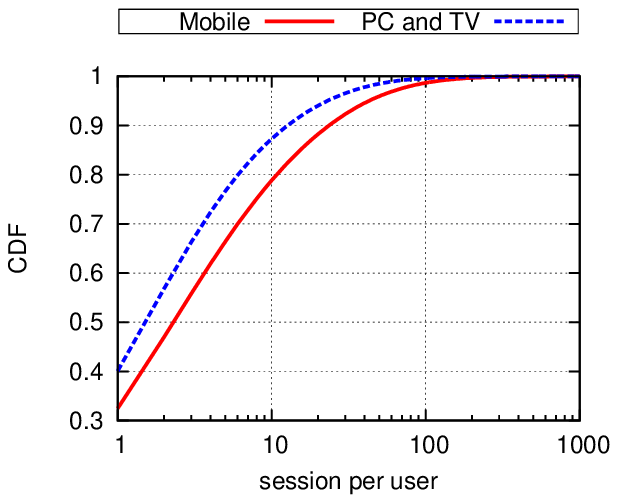}
    }
    \subfloat[Instantaneous load for live content]  {
      \label{fig:live-load}
      \includegraphics[width=0.32\textwidth]{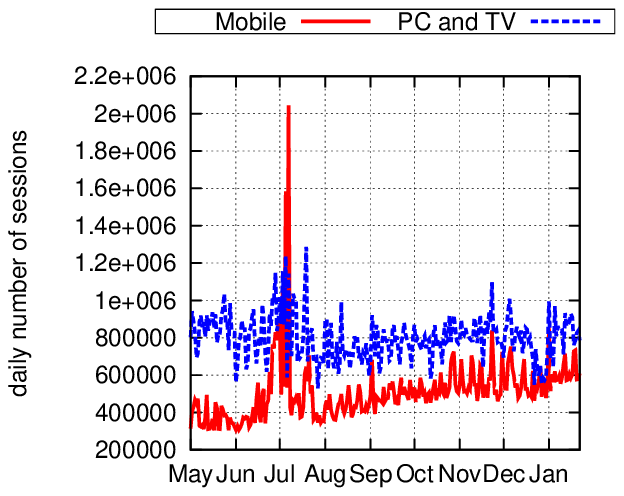}
    }
    \subfloat[Completion ratio by device type]{
      \label{fig:completion-ratio-device}
      \includegraphics[width=0.32\textwidth]{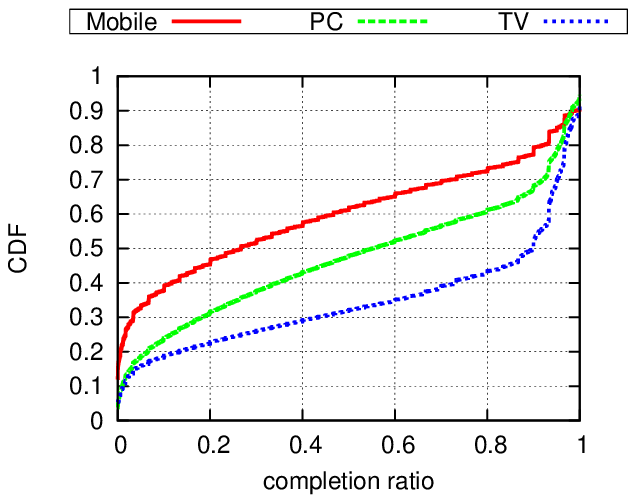}
    }
    \caption{(a) CDF of user activity levels (\#sessions per user) by device type, showing that users of mobile gadgets (smartphones and tablets) are more active than users of fixed devices (PCs and TVs). (b) Mobile users accessing live content can impose significantly higher loads than fixed devices during popular live events such as Wimbledon in July (c) Completion ratio per device type, showing that on-demand sessions are completed less often on mobile devices.
    }\vspace{-6mm}
\end{figure*}


 \subsection{Geographic differences}
 \label{sec:geo}

   To start with, we consider spatial distribution of user activity across the UK\footnote{Note that we only consider users of the fixed-line ISPs for this analysis as the ip-to-location mapping for mobile operators is known to be inaccurate~\cite{triukose2012geolocating}.}. We divide iPlayer sessions based on the region of access, and measure a region's activity level as the average number of sessions of iPlayer users from that region\footnote{$\#users\_from\_region$ accounts for all users who have at least one access from that region.}:
$$
regional\_activity = \frac{\#sessions\_ from\_region}{\#users\_from\_region}
$$
We represent this using a logarithmic scale, and discretise this into five activity levels, from ``low'' to ``high'' in order to plot regional activity levels on a map (Figure~\ref{fig:activity_per_region}). Note that by normalising over the total number of users from a region, this metric accounts for differences in population across regions.

The map shows clear differences across geographic regions: there are very few regions with extremely heavy (olive green) or extremely low (pink) usage. However, the second \emph{highest} point in our usage scale (cyan), is concentrated heavily in the south, in areas around London, whereas the  second \emph{lowest} (purple) levels of access are mainly found in the heavily rural regions of Wales, Scottish highlands and Northern England.


Digging deeper, in \fref{fig:activity_per_region_stats}, we next correlate these differences in usage with regional fixed-line broadband speed statistics, collected by the national regulator Ofcom~\cite{ofcom_geography}. It is clear from the figure that the average regional activity of iPlayer users is highly correlated (Pearson's correlation coefficient of 0.68) with the average Broadband connection bandwidth in the region. This suggests that infrastructural support for high-speed broadband is an important factor that may determine user usage levels for  bandwidth-intensive applications such as BBC iPlayer.

Although the UK government is cognisant of the regional differences in connection speeds through previous studies such as the one from Ofcom~\cite{ofcom_geography}, and has already initiated funding schemes for improving rural broadband access~\cite{uk-broadband} as part of the National Broadband Strategy, we believe this result is important as it provides a quantification of how broadband connection speeds translate to differences at the application level. Having said that, it should be pointed out that \fref{fig:activity_per_region} only shows a \emph{correlation}, and lack of high-speed broadband alone may not fully explain the lower levels of iPlayer usage. Other factors, such as differences in demographics across regions, may also play a role in determining the propensity to use Internet TV streaming applications. 







\subsection{Differences in usage  by device type}
\label{sec:mobile-activity-levels}

We next divide accesses by the type of device used to access iPlayer. We distinguish between PC-like devices (Desktop computers, laptops, etc.), Mobiles (Smartphones and tablets), and Internet-enabled TVs that can be used to access iPlayer. In this subsection, we \emph{do not} distinguish between accesses from mobile ISPs and fixed-line broadband. In particular, with mobile devices, we consider accesses from both mobile and fixed broadband. 

First off, we compare the cumulative distribution of usage levels across these types of devices in \fref{fig:activity-level-gadgets}, again using the number of sessions per user as a metric for activity level. We find that users accessing iPlayer through a mobile device are more active than users from more static devices, with more sessions per user.  

Second, focusing on live events in \fref{fig:live-load}, we find that fixed devices (Desktop computers, laptops, Internet-enabled TVs, etc.) generally tend to have just over a half higher share of traffic than mobile devices. Considering that mobile devices comprise only $\approx20\%$ of accesses, this suggests that mobile devices are being used much more intensively for live usage than other fixed devices. Interestingly, during times of highly popular sports events, such as Wimbledon in June-July, we find that the load from mobile devices for live content can exceed that from all fixed devices combined. 

Finally, we measure the \emph{completion ratio} of sessions from users of different device types. We define completion ratio of a session as follows: Consider a user who is watching a TV programme or content item whose duration is $d_p$. If the duration of the session is $d_u$, then we define the completion ratio $c$ of the session as 
\begin{equation}
c = \frac{d_u}{d_p}
\label{eqn:completion-ratio}
\end{equation}
In \fref{fig:completion-ratio-device}, we consider the cumulative distribution of completion ratio from different device types, and find that mobile users are more impatient, completing much lesser fractions of the shows they watch.

The above differences between mobile and fixed devices could be due to a multitude of reasons: the user demographics of those who access the Internet on their mobile devices could be different and correlate well with heavier usage of Internet in general, and TV streaming in particular. Both the higher number of sessions-per-user as well as the higher usage of live streaming could also simply be a result of the easier access to mobile devices, as their owners can typically be expected to have the device close to them. Another likely possibility is that users of mobile versions of iPlayer are technologically more savvy early adopters who are by definition heavy users. Finally, usage characteristics such as the lower completion ratio could be a result of so-called ``lean forward'' style of content consumption (i.e., using more random access through rewind or fast forwarding, skipping uninteresting bits) with mobile devices~\cite{cui2007personal}, which is more active than the ``lean back'' style of watching traditional TV. 

Regardless of the reasons, these distinct patterns of user behaviour for accesses from mobile devices that imply higher per-device system loads. Although, in absolute numbers, mobile accesses are still in the minority, comprising $\approx20\%$ of all accesses, as mobile devices proliferate and accesses increase over time, the system can expect increased load due to mobile-specific access patterns. We next examine the implications of such differences across fixed-line and mobile providers.






  \begin{figure}
    \centering
\subfloat[Fixed-line Broadband ISPs]{\label{fig:bband-short-term} \includegraphics[width=0.65\columnwidth]{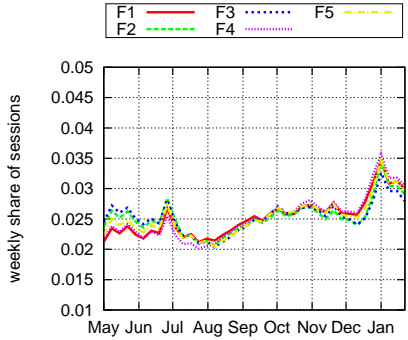}} \\
\subfloat[Mobile providers]{\label{fig:mobile-short-term} \includegraphics[width=0.65\columnwidth]{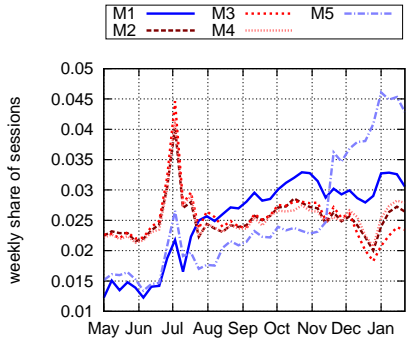}}
\caption{Time-series of user activity across 9 months for (a) fixed-line broadband ISPs and (b) mobile providers. Each line plots the weekly number of accesses for an ISP (normalised by total number of accesses for that ISP). This indicates a relatively stable market for broadband ISPs. Mobile providers with unlimited data plans (blue lines) appear to gain higher shares of user accesses over time than providers with data usage caps (red lines).} 
    \label{fig:time_lapse}
\vspace{-6mm}
  \end{figure}
\begin{figure}
  \centering
  \subfloat[Fixed-line Broadband ISPs]{\label{fig:bband-day}
    \includegraphics[width=0.65\columnwidth]{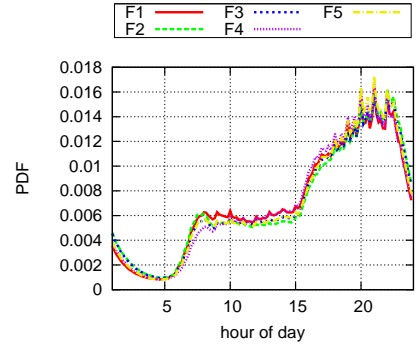}}\\
  \subfloat[Mobile providers]{\label{fig:mobile-day}
    \includegraphics[width=0.65\columnwidth]{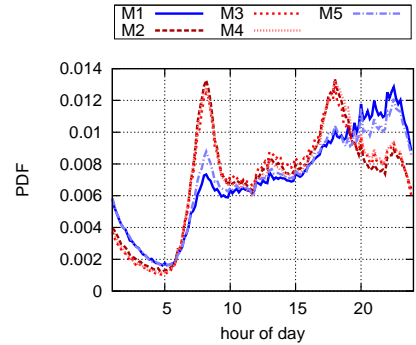}}
  \caption{Diurnal pattern of user activity reveals distinct patterns for fixed-line broadband and mobile providers. (a) Fixed-line broadband providers observe a home-driven access pattern, which peaks during evening hours, when users are expected to be at home, and free to watch TV content. (b) Mobile providers with data usage caps (red lines) observe a commute-driven usage pattern, peaking during morning and evening commute times. In contrast, mobile providers with unlimited data plans (blue lines) see a \emph{superposition} of home- and commute-driven access patterns.}\label{fig:daily_patterns}
\vspace{-6mm}
\end{figure}

\section{ISP-level view of BBC iPlayer usage}
\label{sec:over-time}


Having considered a population level view, we next divide accesses by different types of networks: fixed line broadband and mobile networks. We focus on five representative fixed-broadband and mobile (cellular) network operators.
These are labeled as F1--F5 (fixed-broadband) and M1--M5 (mobile), respectively. First, we analyze the weekly share of sessions from different providers over the entire nine months of the trace to understand the dynamics of the ISP market, and find that while the fixed-line ISP market shares remain stable, the relatively young mobile broadband market can change, with gains for mobile ISPs with unlimited data plans.  Then we examine diurnal patterns of access and find distinct signatures for fixed and mobile ISPs with data usage caps. Interestingly, we find that mobile ISPs with unlimited data plans see a diurnal usage pattern that is a superposition of both kinds of ISPs. Finally, we examine how such data caps and mobile vs.\ fixed connectivity affects the choice of users' session shares across the multiple ISPs from whom they may have access. 





\vspace{-16mm}
\subsection{Temporal dynamics of mobile operators}
\label{sec:9monthview}

First, we consider the number of weekly sessions across the 9 months of our trace, from May 2013 to Jan 2014. Accesses from each of the broadband ISPs and cellular networks by session shares are shown as separate time series in \reffig{fig:time_lapse}. 


In the case of broadband ISPs (\reffig{fig:bband-short-term}), apart from seasonal variations (e.g., increased access during December holidays when users might have more time at home) which affect all the operators considered, we see that the share of user accesses does not change significantly. In contrast, for mobile ISPs (\reffig{fig:mobile-short-term}) we observe two major disruptions: the load of M1 and M5 grows steadily from May 2013 and October 2013, correspondingly. Digging deeper into the cause for this, we observe that only M1, M5 offer unlimited data plans whereas M2, M3 and M4 either did not have unlimited plans or removed any such plans they had during the period we consider. 

\subsection{Differences in diurnal access patterns}

In \reffig{fig:daily_patterns}, we examine diurnal patterns in accesses during January 2014\footnote{This figure uses data from one month, to avoid changes in pricing and other incentives, as well as longer term changes such as daylight-savings time, which affect diurnal patterns slightly. However, we observe similar patterns over each month in our trace.}. To obtain this, each day in the trace was first divided into ten minute epochs, and the number of requests during each epoch was calculated. The figure shows the average number of requests obtained by each ISP, normalised by the total number of requests for that ISP.


Unsurprisingly, for the fixed-line ISPs (\reffig{fig:bband-day}), accesses peak during evening hours,  when users are expected to be at home and presumably, relatively free to watch TV content. An orthogonal pattern is observed for  mobile ISPs with limited-data caps: these providers experience heavy loads during  morning and evening commute times, 
and a smaller peak during lunch time. 
Interestingly, for the unlimited-data mobile networks, the daily access pattern emerges as a superposition of  fixed-line and ``purely'' mobile access, featuring all the different peak hours observed in the two previous cases. 

\subsection{Load share per ISP}
\label{sec:usagecaps}
The previous sections suggest that mobile providers who allow unlimited data usage can drive different, more inclusive traffic patterns incorporating both mobile and at-home usage. This section explores the implications of this observation, and asks how ISP differentiation factors such as data allowances or usage caps  affect the per-user session share of different providers. 

We first define a simple metric to measure relative suitability of a given ISP for a given user. We define \emph{load share} of a  service provider (mobile or fixed-line) for a given user as the fraction of the user's accesses to iPlayer which are from the given service provider. 
A user who has $N$ iPlayer sessions, of which $n_p$ are from provider $p$ obtains a load share of 
\begin{align}
L_p = \frac{n_p}{N}
\end{align}
from that provider. 

Given a user who may have multiple means of obtaining Internet access (e.g., fixed-line broadband at home and work, and/or 3G or other cellular data connectivity through a smartphone or data dongle), this metric is intended to be a gross measure that captures the effect of various different factors (such as price, data limits bandwidth, ubiquitous connectivity, etc.) that may affect the choice of which ISP to access iPlayer from. We do not distinguish the factors, but merely measure the effect of these factors on user choice, in terms of the differential load share across different providers.

\begin{figure}
  \centering
\includegraphics[width=0.75\columnwidth]{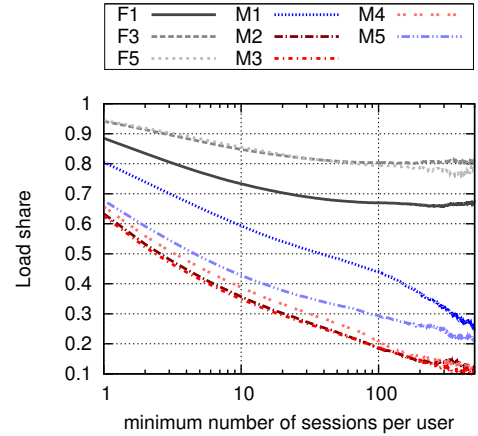}
\caption{Average user load share for different types of providers. Fixed-line broadband has the highest load shares. Mobile providers with usage caps have the least load shares, and their customers who are active users of iPlayer appear to obtain a majority of their accesses from other providers. Mobile providers without usage limits enjoy an intermediate level of load shares.}\vspace{-6mm}
\label{fig:customer_loyality}
\end{figure}

We measure the average user load share of different providers  in \reffig{fig:customer_loyality}, to understand how ISP differentiation factors affect consumer choice. We condition the load share obtained (y-axis) on the minimum activity level of users in terms of number of sessions (x-axis), since highly active users may also be sophisticated online citizens with multiple forms of Internet connectivity and therefore decreased load share. Validating this assumption, \reffig{fig:customer_loyality} shows that average load share to any single provider decreases amongst customers as activity increases. As may be expected, it is seen that customers of the fixed-line broadband ISPs have the largest fraction of sessions from a sole internet access point. In contrast, mobile ISPs with data caps  have the least load shares. It appears that the limited-data caps force customers to look elsewhere for cheaper or more suitable Internet access options to satisfy their demands of the high bandwidth data. Mobile users with unlimited data plans in general spend a higher share of accesses with their mobile operator than do users with limited plans, suggesting that unlimited plans can be a good incentive for iPlayer users to use a single ISP. 

\section{Alleviating the limitations of data-caps with Cross-ISP interactions}
\label{sec:crossisp}
The previous section suggested that data caps can have a huge impact on the relative suitability of an ISP for a user who wishes to use an application like iPlayer. However, data caps are a hard economic necessity for many mobile providers (and users as well, since unlimited plans may have different pricing structure). In this section, we explore how the needs of mobile users can be supported effectively in the face of data caps.

To address this issue, we observe that when the mobile ISP cannot fully satisfy a user's needs, her load share will be less than 1, which immediately implies that the user roams across different ISPs. We propose to consider how such cross-ISP interactions of users can be used to support user needs, and suggest two strategies. One is based on product bundling discounts, already being practised by some mobile providers in the UK. The second, based on exploiting the regularity of user accesses, predictively pre-fetches content through a fixed-line broadband connection before it is used on a mobile network, and thereby removes such access from being subject to data caps. Note that both strategies are complementary to each other and may be used independently or in conjunction.

\subsection{Product bundling}
\label{sec:interisp}
This strategy is based on the insight that it is easier or less expensive to provide high bandwidth access through fixed-line broadband than it is to do so using a cellular architecture. Therefore, one solution to the limitations imposed by a mobile provider's data cap is to provide the user with fixed-line broadband connectivity at home. Whilst the consumer can separately buy mobile data connectivity from one provider and fixed-line broadband from another provider, some broadband and mobile ISPs provide product bundles to make the combination of the two ISPs more attractive to consumers. Similarly, some cellular operators operate their own fixed-line broadband, and some broadband providers may also be Mobile Virtual  Network Operators. In this section, we measure the effectiveness of this strategy in capturing a greater share of user loads.

Consider two providers $i$ and $j$ with  shares $\rho_i$ and $\rho_j$ respectively of all user accesses. A randomly chosen user has accesses from provider $i$ with probability $\rho_i$,  provider $j$ with probability $\rho_j$ and from both providers with a probability $\rho_{i,j} = \rho_i \rho_j$ (assuming that a user's choice between operator $i$ and operator $j$ is independent). If in our trace we observe a fraction $\overline{\rho_{i,j}}$ of users with accesses from both $i$ and $j$, then we define an inter-service provider attractiveness coefficient  $\gamma_{i,j}$ for $i$ and $j$ as: 
\begin{align}
\gamma_{i,j} = \log{[\frac{\overline{\rho_{i,j}}}{\rho_{i,j}}]} = \log{[\frac{\overline{\rho_{i,j}}}{\rho_i  \rho_j}]}
\end{align}

We say that providers $i$ and $j$ attract each other if $\gamma_{i,j} > 0$ (i.e., more users are in both providers than can be expected by chance). $i$ and $j$ can be considered to ``repel'' each other otherwise (i.e., fewer users are in both providers than can be expected by chance). We next compute the Inter-Service Provider attractiveness for each pair (named V1 - V30) of providers we consider (M1--M5 and F1--F5\footnote{Note that one of the fixed-line ISPs has been separated out into two ISPs reflecting the situation  prior to a recent merger. This creates 30 virtual combinations rather than the expected 25, but allows us to better see the attractiveness, as only one of the two ISPs which merged had offered product bundling to its customers.}). 


\begin{figure}
  \centering
  \includegraphics[width=0.75\columnwidth]{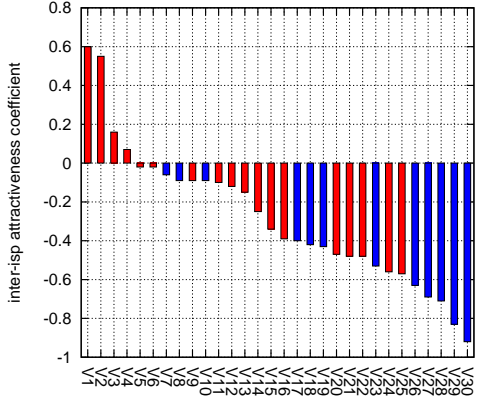}
  \caption{Inter-ISP attractiveness coefficients, measuring whether the number of common users between a fixed and mobile provider is greater than chance (attractiveness $>0$), or not (attractiveness $<0$). Each bar represents a combination of a fixed line provider with a mobile provider with unlimited data plans (blue) or limited data caps (red). The bundles with positive attractiveness coefficients (V1-V4) correspond to an existing {product bundle} when home broadband is offered with discounts and special deals for mobile customers that make it attractive for customers to choose both providers together.} 
\vspace{-8mm}
  \label{fig:cross_isp_ties}
\end{figure}

Considering the attractiveness between mobile and fixed providers reveals an interesting insight (Figure~\ref{fig:cross_isp_ties}): Mobile providers with unlimited data plans (blue bars) appear to repel other fixed-line ISPs, whereas operators in bundles $V1 - V4$ appear to have more users in common than can be expected by chance. Digging deeper into the cause of the latter we found that each of these links corresonds to an existing product bundle on the market when a home broadband provider offers  discounts and special deals for the corresponding mobile operator that make it attractive for customers to choose both providers as a product bundle. 
We believe these examples provide some confirmatory evidence of the efficacy of ``product bundling'' as a pricing strategy to attract and retain customers across fixed and mobile operators. An important implication of well-defined pairing of a mobile operator with a fixed-line broadband operator is that some of the mobile access can be offloaded by pre-fetching using the fixed-line connection, as discussed below.

\subsection{Speculative prefetching}
The second strategy we consider is one that can be implemented directly by a content provider, without ISP assistance. The basic idea is to offload the mobile traffic to a fixed-line broadband operator by using speculative pre-fetching strategies such as the ones extensively explored in previous literature~\cite{nencioni2013understanding,siris2013enhancing, lu2013optimizing, kamaraju2013novel, hoque2013using, wang2013cloud, kaup2013optimizing}.  Alternately, a pair of fixed-line and mobile providers can collaborate with each other to aggresively pre-fetch content expected to be accessed by the user over a mobile network, and push it to a device-local cache  controlled by the mobile network operator (e.g., pushed to reserved storage on the phone). In this subsection, we explore the feasibility of such an approach.

\begin{figure}
\centering  
  \subfloat[Regularity of arrivals]{\centering
    \label{fig:dev-returntime}
    \includegraphics[width=0.60\columnwidth]{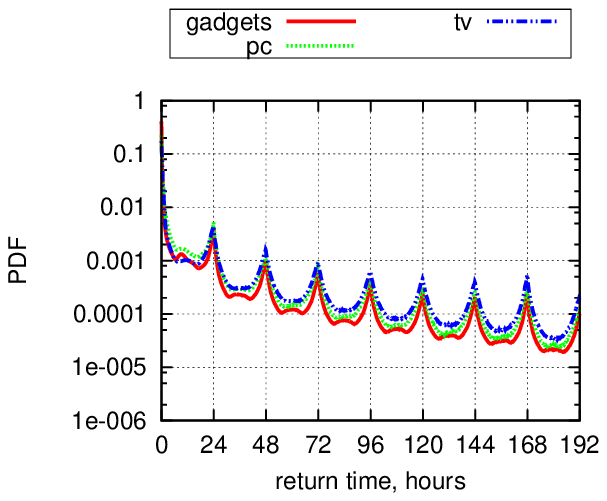}\hspace{5mm}
  }\\
  \subfloat[Completion across different ISPs]{\centering
    \label{fig:isp-completion}
    \includegraphics[width=0.6\columnwidth]{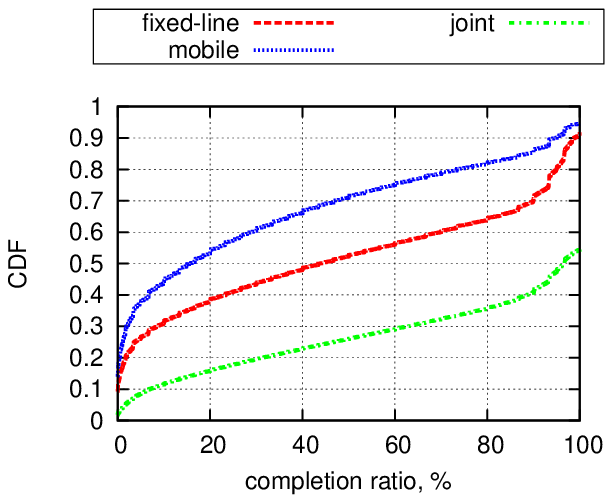}
  }
  \caption{Regularity of arrivals (a): Distribution of time intervals between consecutive sessions of each user: users tend to re-visit iPlayer after time intervals proportional to 24 hours. Distribution of completion ratios split across ISPs (b): users of mobile ISPs have lower completion ratios than users of fixed-line internet.}\vspace{-6mm}
  \label{fig:completion_ratio}
  \end{figure}

First, we study how people access content across mobile and broadband operators. To this end, we first measure how often users return back to iPlayer. \fref{fig:dev-returntime} shows that users are more likely to return back to iPlayer after time intervals proportional to $24$ hours. This periodicity could be exploited to speculatively pre-fetch content whilst on a fixed connection, before the next access on a mobile connection happens. 
 
We also measure the fractions of content which users watch from mobile and fixed-line ISPs and notice that mobile users have lower completion ratios (as defined in Equation~\eqref{eqn:completion-ratio}) with respect to fixed-line users (\fref{fig:isp-completion})\footnote{This result is distinct from \fref{fig:completion-ratio-device}, which studies completion across different device types across any kind of network. Here we study completion ratio of any device, across fixed and mobile ISPs}. Consistent with the patterns observed with return times (\fref{fig:dev-returntime}) and access times (\fref{fig:daily_patterns}), this arises as mobile users watch episodes partially during morning or evening commutes, the hours of their highest activity. 

Interestingly, \fref{fig:isp-completion} also shows that completion ratio increases significantly if we join together sessions from the same user to the same content item. We consider on-demand sessions which are started from a fixed networks and finished when on the move, or vise versa. On a random sample of 1M users, we observed that approximately 52\% of such joint sessions are started on the fixed network. Thus, when the user is still on the fixed-line ISP, content can be aggressively pre-fetched beyond the current watch point, and cached on the mobile device. When the user accesses the rest of the show from a mobile network, it can replayed from cache, thereby preventing the session from counting against the user's data caps on the mobile network. Of the remaining 48\% of sessions, which are all started on the mobile network, 23\% are for new episodes of shows whose previous episodes have also been watched by the user. For example, a user who has previously watched  ``Dr.\ Who, Episode $j$'', either on fixed or mobile connection, might watch some episode, ``Dr.\ Who, Episode $j+k$'' ($k>0$), while on the move. By exploiting previous watch history, new episodes of ``Dr.\ Who'' may be speculatively pre-fetched onto  the mobile device,  thereby offloading a future access from a mobile connection. Clearly, such speculative strategies also have to consider how to manage the limited amount of storage which might be available on the mobile device. A full study of this issue and an implementable solution is beyond the scope of this study. However the extensive prior work on speculative pre-fetching ~\cite{nencioni2013understanding,siris2013enhancing} has considered similar issues.

\subsection{Limitations to mobile-fixed interactions}
	\begin{figure}
    \centering
      \subfloat[Correlation of content popularity across ISPs]{
\hspace{9mm}\includegraphics[width=0.6\columnwidth]{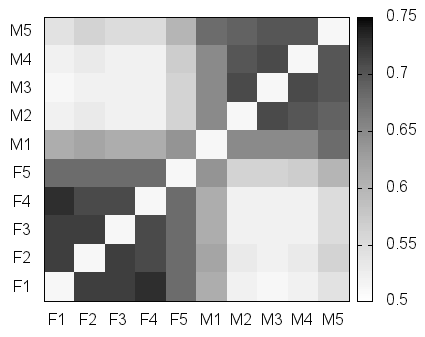}\vspace{-5mm}
}\\
			\subfloat[Share of access per content genres]{
        \includegraphics[width=0.6\columnwidth]{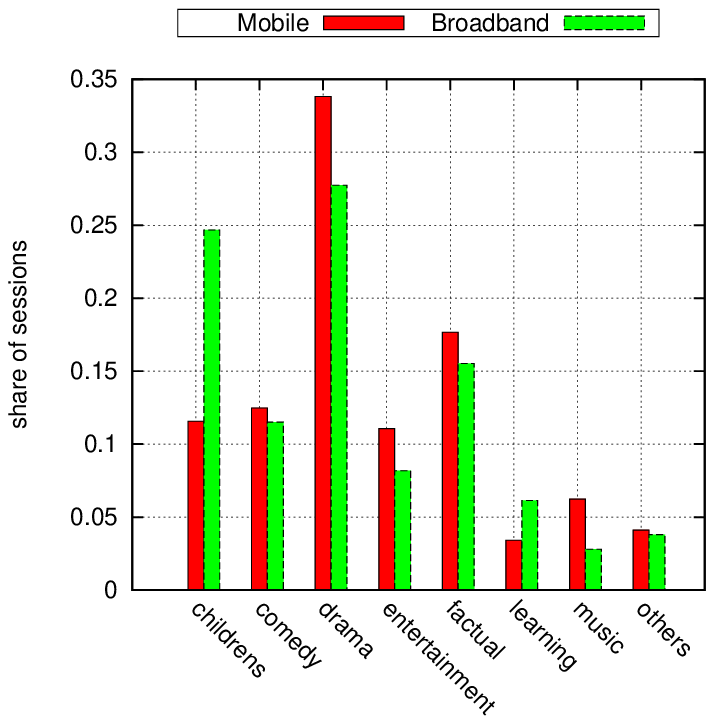}}
    \caption{Comparison of the content items watched by users of mobile and fixed-line ISPs shows that: a) there is a high  Spearman's correlation of content popularity between mobile ISPs, and similarly between Fixed ISPs, but correlation is low between a mobile ISP and a fixed ISP. b) the share of access for children content, highly popular among fixed-line users, drops dramatically for mobile users.}\vspace{-8mm}
    \label{fig:content_popularity_correlation}
  \end{figure}
	
Finally, we analyze what people watch across different kinds of ISPs. We split iPlayer accesses by the ISPs they come from and rank content items according to their number of accesses in each network. Then, we compute Spearman's correlation coefficients between the content popularity rankings for each pair of networks. Figure~\ref{fig:content_popularity_correlation}a shows a correlation matrix for every pair of operators. Notice that the correlation is high for the ISPs within the mobile and fixed-line groups and is significantly lower for pairs of ISPs from different groups. Additionally, we note from Figure~\ref{fig:content_popularity_correlation}b that children and learning content which accounts for a giant ($\approx$25\%) fraction of fixed-line accesses drops dramatically for mobile users: evidently, adult users dominate for mobile broadband access. 


 We envisage that such discrepancies in content popularity across different networks is a limitation for mobile-fixed network co-operation which should be accounted for in the design of prefetching mechanisms. On the other hand, it opens a door for co-operation between mobile operators (or between fixed-line ISPs) with similar content access patterns. This can potentially reduce maintenance and infrastructure upgrade costs by installing content caches commonly shared across all co-operating ISPs. We note that these ideas are in line with a general trend in the UK's mobile market where two rival operators have  recently announced a deal to share 4G infrastructure~\cite{4g_cooperation}.





\section{Discussion and conclusion}


We offer, to the best of our knowledge, the first \emph{population-level picture of a  nation's Internet usage}, that covers an estimated 50\% of UK's online population, comparing and contrasting the relative markets for fixed-line and mobile (cellular) broadband access, from the perspective of one application -- online streaming. 

Our study found evidence that connection speeds of different regions of the UK correlate with the regional activity levels, the average number of sessions per user, underscoring the importance of speed. We also found that mobile devices can have disproportionately large share of live accesses, especially during popular sports events such as Wimbledon. Mobile devices also complete smaller fractions of a programme in a single session. 

We then surveyed 5 representative mobile and fixed broadband ISPs.  The more mature broadband ISP market remained relatively stable over the nine months of our trace, but the mobile market proved to be dynamic: we observed an increase in session shares for mobile providers who offer unlimited data plans. We observed distinct differences in the diurnal access patterns across different kinds of providers: sessions from fixed-line broadband ISPs peak during evening hours, when users are expected to be at home; mobile providers with data caps contribute sessions during commute times; and mobile providers with unlimited data plans show a \emph{superposition} of both patterns, with peaks during commute-time \emph{and} during evening times. 

To measure to what extent different providers capture the market of each user, we defined a simple measure called load share that captures the per-user session share that goes to a given provider, and showed that fixed-line ISPs are able to capture a greater share of user sessions in comparison with mobile providers. Similarly, mobile providers with unlimited data obtain a greater load share than mobile providers with limited data. However, some of the providers with limited data are able to capture a greater share of their users' sessions by introducing discounts and tie-ups with preferred fixed-line broadband providers, creating an ``inter-ISP attractiveness'' that helps capture a greater share of user accesses. 
Finally, we studied how the roaming of users across multiple ISPs can be used to advantage by speculatively pre-fetching content over fixed-line connections and offloading it from mobile access. 

Our study was enabled by the large coverage that iPlayer has of the UK popoulation online. This is by no means a unique phenomenon -- indeed, the ``winner take all'' nature of modern Web applications mean that other Internet giants, such as Google, Facebook or Akamai may have similar views of the online population, perhaps also on a global scale. Exploiting this for other Internet-scale measurements may be an interesting avenue for future research.

\bibliographystyle{acm}
\bibliography{section/biblio}

\begin{thebibliography}{10}

\bibitem{akamai-soti}
{\sc Akamai}.
\newblock State of the internet, Q4 2013.

\bibitem{bbc_performance_pack_2014_march}
{\sc BBC}.
\newblock Bbc iplayer performance pack march 2014.
\newblock \url{http://goo.gl/f3stgR}.

\bibitem{cui2007personal}
{\sc Cui, Y., Chipchase, J., and Jung, Y.}
\newblock Personal tv: A qualitative study of mobile tv users.
\newblock In {\em Interactive TV: A shared experience}. Springer, 2007,
  pp.~195--204.

\bibitem{uk-broadband}
{\sc {Department for Culture Media, \& Sport}, {Parliamentary Under Secretary
  of State for Culture, Communications and Creative Industries}, and {The Rt.\
  Hon.\ Sajid Javid MP}}.
\newblock Stimulating private sector investment to achieve a transformation in
  broadband in the uk by 2015.
\newblock Available from
  \url{https://www.gov.uk/government/policies/transforming-uk-broadband},
  February 2013.

\bibitem{dischinger2007characterizing}
{\sc Dischinger, M., Haeberlen, A., Gummadi, K.~P., and Saroiu, S.}
\newblock Characterizing residential broadband networks.
\newblock In {\em Proceedings of the IMC\/} (2007), pp.~43--56.

\bibitem{heidemann2008census}
{\sc Heidemann, J., Pradkin, Y., Govindan, R., Papadopoulos, C., Bartlett, G.,
  and Bannister, J.}
\newblock Census and survey of the visible internet.
\newblock In {\em Proceedings of the 8th ACM SIGCOMM IMC\/} (2008), ACM,
  pp.~169--182.

\bibitem{hoque2013using}
{\sc Hoque, M.~A., Siekkinen, M., and Nurminen, J.~K.}
\newblock Using crowd-sourced viewing statistics to save energy in wireless
  video streaming.
\newblock In {\em Proceedings of the 19th MobiCom\/} (2013), ACM, pp.~377--388.

\bibitem{huang2013depth}
{\sc Huang, J., Qian, F., Guo, Y., Zhou, Y., Xu, Q., Mao, Z.~M., Sen, S., and
  Spatscheck, O.}
\newblock An in-depth study of lte: Effect of network protocol and application
  behavior on performance.
\newblock In {\em Proceedings of the ACM SIGCOMM 2013\/} (2013), ACM,
  pp.~363--374.

\bibitem{kamaraju2013novel}
{\sc Kamaraju, P., Lungaro, P., and Segall, Z.}
\newblock A novel paradigm for context-aware content pre-fetching in mobile
  networks.
\newblock In {\em WCNC, 2013 IEEE\/} (2013), IEEE, pp.~4534--4539.

\bibitem{karagiannis2005blinc}
{\sc Karagiannis, T., Papagiannaki, K., and Faloutsos, M.}
\newblock Blinc: multilevel traffic classification in the dark.
\newblock In {\em ACM SIGCOMM Computer Communication Review\/} (2005), vol.~35,
  ACM, pp.~229--240.

\bibitem{kaup2013optimizing}
{\sc Kaup, F., and Hausheer, D.}
\newblock Optimizing energy consumption and qoe on mobile devices.
\newblock In {\em ICNP\/} (2013), pp.~1--3.

\bibitem{kreibich2010netalyzr}
{\sc Kreibich, C., Weaver, N., Nechaev, B., and Paxson, V.}
\newblock Netalyzr: illuminating the edge network.
\newblock In {\em Proceedings of the 10th ACM SIGCOMM IMC\/} (2010), ACM,
  pp.~246--259.

\bibitem{lakshminarayanan2003some}
{\sc Lakshminarayanan, K., and Padmanabhan, V.~N.}
\newblock Some findings on the network performance of broadband hosts.
\newblock In {\em Proceedings of the 3rd ACM SIGCOMM IMC\/} (2003), ACM,
  pp.~45--50.

\bibitem{li2012watching}
{\sc Li, Z., Lin, J., Akodjenou, M.-I., Xie, G., Kaafar, M.~A., Jin, Y., and
  Peng, G.}
\newblock Watching videos from everywhere: a study of the pptv mobile vod
  system.
\newblock In {\em Proceedings of the 2012 ACM IMC\/} (2012), ACM, pp.~185--198.

\bibitem{lu2013optimizing}
{\sc Lu, Z., and de~Veciana, G.}
\newblock Optimizing stored video delivery for mobile networks: The value of
  knowing the future.
\newblock In {\em INFOCOM, 2013 Proceedings IEEE\/} (2013), IEEE,
  pp.~2706--2714.

\bibitem{maier2009dominant}
{\sc Maier, G., Feldmann, A., Paxson, V., and Allman, M.}
\newblock On dominant characteristics of residential broadband internet
  traffic.
\newblock In {\em Proceedings of the 9th ACM SIGCOMM IMC\/} (2009), ACM,
  pp.~90--102.

\bibitem{moore2005toward}
{\sc Moore, A.~W., and Papagiannaki, K.}
\newblock Toward the accurate identification of network applications.
\newblock In {\em Passive and Active Network Measurement}. Springer, 2005,
  pp.~41--54.

\bibitem{nencioni2013understanding}
{\sc Nencioni, G., Sastry, N., Chandaria, J., and Crowcroft, J.}
\newblock Understanding and decreasing the network footprint of catch-up tv.
\newblock In {\em Proceedings of the WWW\/} (2013), pp.~965--976.

\bibitem{ofcom}
{\sc Ofcom}.
\newblock Communications market report july, 2012.
\newblock Available from \url{http://goo.gl/cSQGvL}.

\bibitem{ofcom_geography}
{\sc Ofcom}.
\newblock Uk fixed broadband map 2013.
\newblock Available from \url{http://maps.ofcom.org.uk/broadband/}, 2013.

\bibitem{sanchez2013dasu}
{\sc S{\'a}nchez, M.~A., Otto, J.~S., Bischof, Z.~S., Choffnes, D.~R.,
  Bustamante, F.~E., Krishnamurthy, B., and Willinger, W.}
\newblock Dasu: Pushing experiments to the internet's edge.
\newblock In {\em USENIX NSDI\/} (2013), vol.~2013, pp.~487--499.

\bibitem{sandvine}
{\sc Sandvine}.
\newblock Global internet phenomena report, 2h 2012, May 2012.

\bibitem{netiathome}
{\sc Simpson~Jr, C.~R., and Riley, G.~F.}
\newblock Neti@ home: A distributed approach to collecting end-to-end network
  performance measurements.
\newblock In {\em Passive and Active Network Measurement}. Springer, 2004,
  pp.~168--174.

\bibitem{siris2013enhancing}
{\sc Siris, V.~A., and Kalyvas, D.}
\newblock Enhancing mobile data offloading with mobility prediction and
  prefetching.
\newblock {\em ACM SIGMOBILE Mobile Computing and Communications Review 17}, 1
  (2013), 22--29.

\bibitem{sundaresan2011broadband}
{\sc Sundaresan, S., De~Donato, W., Feamster, N., Teixeira, R., Crawford, S.,
  and Pescap{\`e}, A.}
\newblock Broadband internet performance: a view from the gateway.
\newblock In {\em ACM SIGCOMM Computer Communication Review\/} (2011), vol.~41,
  ACM, pp.~134--145.

\bibitem{4g_cooperation}
{\sc TheInquirer.net}.
\newblock Rival operators ee and three strike a deal to share 4g
  infrastructure.
\newblock Available from \url{http://goo.gl/13AgIs}, 2014.

\bibitem{triukose2012geolocating}
{\sc Triukose, S., Ardon, S., Mahanti, A., and Seth, A.}
\newblock Geolocating ip addresses in cellular data networks.
\newblock In {\em Passive and Active Measurement\/} (2012), Springer,
  pp.~158--167.

\bibitem{wang2013cloud}
{\sc Wang, X., Kwon, T.~T., Choi, Y., Wang, H., and Liu, J.}
\newblock Cloud-assisted adaptive video streaming and social-aware video
  prefetching for mobile users.
\newblock {\em Wireless Communications, IEEE 20}, 3 (2013).

\bibitem{xu2005profiling}
{\sc Xu, K., Zhang, Z.-L., and Bhattacharyya, S.}
\newblock Profiling internet backbone traffic: behavior models and
  applications.
\newblock In {\em ACM SIGCOMM Computer Communication Review\/} (2005), vol.~35,
  ACM, pp.~169--180.

\bibitem{xu2011cellular}
{\sc Xu, Q., Huang, J., Wang, Z., Qian, F., Gerber, A., and Mao, Z.~M.}
\newblock Cellular data network infrastructure characterization and implication
  on mobile content placement.
\newblock In {\em Proceedings of the ACM SIGMETRICS\/} (2011), ACM,
  pp.~317--328.

\bibitem{zhu2011characterizing}
{\sc Zhu, Z., Cao, G., Keralapura, R., and Nucci, A.}
\newblock Characterizing data services in a 3g network: Usage, mobility and
  access issues.
\newblock In {\em Proceedings of the IEEE ICC 2011\/} (2011), IEEE, pp.~1--6.

\end{thebibliography}
 
\end{document}